\documentclass[a4paper,twocolumn,11pt,accepted=2024-01-30]{quantumarticle}
\pdfoutput=1
\usepackage[numbers,sort&compress]{natbib}
\usepackage{amsmath}
\usepackage{amsfonts}
\newcommand{\onlinecite}[1]{\cite{#1}}
\renewcommand{\appendix}{\onecolumn\newpage}

\usepackage{graphicx}
\usepackage{dcolumn}
\usepackage{bm}
\usepackage{hyperref}
\usepackage{braket}
\usepackage{enumitem}
\usepackage{xcolor}
\usepackage{transparent}
\usepackage[export]{adjustbox}
\usepackage{tikz}
\usetikzlibrary{positioning}
\graphicspath{{images/}}

\renewcommand{\Re}{\operatorname{Re}}
\renewcommand{\Im}{\operatorname{Im}}

\usepackage{pifont}
\newcommand{\cmark}{\ding{51}}
\newcommand{\xmark}{\ding{55}}

\newcommand{\toadd}[1]{{#1}}

\begin{document}


\title{Impact of conditional modelling for a universal autoregressive quantum state}


\author{Massimo Bortone}
    \email{massimo.bortone@kcl.ac.uk}
\author{Yannic Rath}%
    \email{yannic.rath@kcl.ac.uk}
\author{George~H.~Booth}%
    \email{george.booth@kcl.ac.uk}
\affiliation{%
    Department of Physics, King’s College London, Strand, London WC2R 2LS, United Kingdom
}%


\begin{abstract}
    We present a generalized framework to adapt universal quantum state approximators, enabling them to satisfy rigorous normalization and autoregressive properties.
    We also introduce filters as analogues to convolutional layers in neural networks to incorporate translationally symmetrized correlations in arbitrary quantum states.
    By applying this framework to the Gaussian process state, we enforce autoregressive and/or filter properties, analyzing the impact of the resulting inductive biases on variational flexibility, symmetries, and conserved quantities.
    In doing so we bring together different autoregressive states under a unified framework for machine learning-inspired ans\"atze.
    Our results provide insights into how the autoregressive construction influences the ability of a variational model to describe correlations in spin and fermionic lattice models, as well as {\em ab initio} electronic structure problems where the choice of representation affects accuracy.
    We conclude that, while enabling efficient and direct sampling, thus avoiding autocorrelation and loss of ergodicity issues in Metropolis sampling, the autoregressive construction materially constrains the expressivity of the model in many systems.
\end{abstract}

\maketitle


\section{Introduction}
The quantum many-body problem is a keystone challenge in the description of quantum matter from nuclei to materials and many more fields besides.
Its formal solution scales exponential with number of interacting particles, but recent developments in neural and tensor network representations have made significant advances in defining compact and expressive approximations to the many-body wave function.
This has allowed for accurate solutions to many complex quantum systems in condensed matter physics~\cite{nomuraDiracTypeNodalSpin2021,rothHighaccuracyVariationalMonte2023,chenEfficientOptimizationDeep2023}, quantum chemistry~\cite{pfauInitioSolutionManyelectron2020,hermannInitioQuantumChemistry2023,hermannDeepneuralnetworkSolutionElectronic2020,chooFermionicNeuralnetworkStates2020} and beyond~\cite{jonssonNeuralnetworkStatesClassical2018,torlaiNeuralnetworkQuantumState2018,medvidovicClassicalVariationalSimulation2021,lovatoHiddennucleonsNeuralnetworkQuantum2022}.
Neural Quantum States (NQS) use neural networks as polynomially compact models of the wave function, and are variationally optimized via stochastic sampling of expectation values.
Many NQS architectures have been investigated in recent years, starting with the Restricted Boltzmann Machine (RBM)~\cite{carleoSolvingQuantumManybody2017}.
However, it has been shown that a state parameterization inspired by kernel models rather than neural networks can also be used to achieve a similar level of accuracy and flexibility with a simpler functional form, derived straightforwardly from well-defined physical arguments.
These `Gaussian process states' (GPS)~\cite{glielmoGaussianProcessStates2020,rathBayesianInferenceFramework2020,rathQuantumGaussianProcess2022}, as well as other related kernel models~\cite{giulianiLearningGroundStates2023a}, have been shown to efficiently and compactly represent a large class of low-energy quantum states to high-accuracy\toadd{, and can be considered as a member of the broader `NQS' family of parameterized quantum states}.


Given the many alternative functional forms that the different machine learning-inspired architectures can imply, a significant advance from the initial demonstration of the NQS in quantum systems has been the refining of particular models based on enforcing desired properties.
Motivated by the performance of deep convolutional neural networks (CNN) in the field of computer vision~\cite{rawatDeepConvolutionalNeural2017}, wave function models that incorporate many layers of translationally-invariant convolutional filters to efficiently learn local correlated features have been proposed and applied to find ground states of frustrated quantum spin systems~\cite{chooTwodimensionalFrustratedText2019}.
The recent success of {\em autoregressive} (AR) generative models in machine learning (ML)~\cite{bond-taylorDeepGenerativeModelling2022} has also captured the attention of physicists interested in the quantum many-body problem, leading to the development of autoregressive quantum states (ARQS) that enforce a strictly normalized state from which configurations can be directly sampled without Metropolis Monte Carlo, autocorrelation times or loss of ergodicity.

In this context, Sharir et al.~\cite{sharirDeepAutoregressiveModels2020} were the first to propose an adaption of PixelCNN~\cite{vandenoordConditionalImageGeneration2016} (an autoregressive masked convolutional neural network for image generation) to the quantum many-body problem and applied it to find the ground state of two-dimensional transverse-field Ising and antiferromagnetic Heisenberg systems.
Other ML architectures such as recurrent neural networks (RNN)~\cite{hibat-allahRecurrentNeuralNetwork2020,hibat-allahSupplementingRecurrentNeural2022} and transformer architectures~\cite{zhangTransformerQuantumState2023} have also been proposed as models for ARQS, yielding convincing results about their ability to represent ground states of lattice systems with different geometries and to compute accurate entanglement entropies in systems with topological order~\cite{hibat-allahInvestigatingTopologicalOrder2023a}.
Hybrid models that combine the expressivity of autoregressive architectures from the deep learning literature with the physical inductive bias of tensor networks have also been proposed~\cite{wuTensornetworkQuantumStates2023,chenANTNBridgingAutoregressive2023}.
Going beyond quantum spin lattice systems, extensions of ARQS based on deep feed-forward neural networks have been applied to the {\em ab initio} electronic structure problem in quantum chemistry, demonstrating good accuracy up to 30 spin-orbitals~\cite{barrettAutoregressiveNeuralnetworkWavefunctions2022}. 

At the core of any autoregressive model is the application of the product rule of probability to factorize a joint-probability distribution of $N$ random variables into a causal product of probability distributions, one for each variable, conditioned on the realization of previous variables.
This modelling approach can be extended to quantum states, yielding explicitly normalized autoregressive ansätze from which independent configurations can be sampled directly via a sequential process.
This ability is of particular interest in the context of Monte Carlo, where the computation of expectation values via autocorrelated stochastic processes such as Metropolis sampling can lead to loss of ergodicity or long sampling times \cite{sharirDeepAutoregressiveModels2020}.

In this work, we describe \toadd{a procedure} to adapt {\em general} quantum states into an autoregressive form, as well as introduce filters to improve the parameter scaling, enforce translational symmetry and \toadd{exploit} locality of correlation features \toadd{in the model.
We specifically apply these adaptations to the GPS model to introduce autoregressive and filter variants of this model. These procedures will however also allow for autoregressive and filter/convolutional adaptations of other wave function ans\"atze. Since the GPS model has a simpler analytical form for the `parent' state compared to many NQS architectures (while sharing its properties as a universal approximator and similar compact expressibility of many quantum states), the resulting autoregressive GPS ansatz is also particularly simple to analyze, while sharing many properties of more complex AR models.}

\toadd{This allows us to disentangle the impact of the different conditions required for an AR state on the flexibility of the resulting model. In particular, it is not clear in general how enforcing autoregressive properties affects the resulting expressibility of the state compared to its parent model.
While the advantages of direct sampling of configurations from autoregressive ans\"atze has been well demonstrated (though its impact is system-dependent)~\cite{zhaoOvercomingBarriersScalability2021,zhaoScalableNeuralQuantum2023}, it has been less clear how the different conditions required for it (such as the masking, the normalization, and the more limited choice of symmetrization) reduce the overall variational freedom afforded by the state. We will investigate these questions, by }
directly comparing the autoregressive state to its parent (unnormalized) GPS, and discussing the advantages and otherwise in the choice of AR models in general. \toadd{We find the} normalization of the conditionals to be the dominant factor in the expressibility of these \toadd{AR-adapted GPS} states for an unfrustrated 2D spin lattice. 
We consider results on spin models, fermionic lattices and {\em ab initio} systems, considering further the impact of sign structure, choice of representation and numerical expediency. \toadd{While we only provide numerical evidence for the impact of these adaptations for the GPS parent model, we would hypothesise that these qualitative conclusions are more broadly valid for autoregressive NQS architectures due to the constraints that this necessarily imposes, and can potentially be used as a guiding principle for the design of future AR models.}

\section{A framework for compact many-body wave functions}\label{sec:uni-framework-wf}
\subsection{Quantum states as product of correlation functions}\label{subsec:size-ext-prod-sep}

The many-body quantum state of a given system consisting of $N$ modes, each represented by a local Fock space of $D$ local states as $x_i\in\{0,\dots,D-1\}$, is fully described by a set of $D^N$ amplitudes $\psi_{x_1,\dots,x_N}$ and basis configurations $\ket{\mathbf{x}}$, i.e.
\begin{equation}\label{eq:qmb-wf}
    \ket{\psi} = \sum_{\mathbf{x}} \psi_{x_1,\dots,x_N}\ket{\mathbf{x}},
\end{equation}
where $\mathbf{x}=(x_1,\dots,x_N)$ is a string representing the local states of each mode in the configuration $\ket{\mathbf{x}}$.
This presents a challenging problem, since the number of amplitudes grows exponentially with system size (number of modes, sites in a lattice or number of orbitals in {\em ab initio} systems).

The variational Monte Carlo (VMC) approach circumvents this by replacing the structureless tensor $\psi_{x_1,\dots,x_N}$ in Eq. \ref{eq:qmb-wf} with a model that can be efficiently evaluated at any configuration, $\psi_{\theta}: \{0,\dots,D-1\}^N \rightarrow \mathbb{C}$, parameterized by a vector $\theta$ of size $\mathcal{O}(\text{poly}(N))$.
This compact representation of the wave function then enables the estimation of expectation values of operators $\hat{O}$ via stochastic evaluation, as
\begin{align}
    \langle\hat{O}\rangle &= \braket{\psi_\theta|\hat{O}|\psi_\theta} \\
    &= \sum_{\mathbf{x}}|\psi_\theta(\mathbf{x})|^2\sum_{\mathbf{x}'}O_{\mathbf{x}\mathbf{x}'}\frac{\psi_\theta(\mathbf{x}')}{\psi_\theta(\mathbf{x})} \\
    &= \mathbf{E}_{\mathbf{x}\sim p_\theta}\left[O_{loc}(\mathbf{x})\right] \label{eq:exp-val},
\end{align}
where $p_\theta(\mathbf{x})=|\psi_\theta(\mathbf{x})|^2$ is the Born probability of $\mathbf{x}$ and $O_{loc}(\mathbf{x})=\sum_{\mathbf{x}'}O_{\mathbf{x}\mathbf{x}'}\psi_\theta(\mathbf{x}')/\psi_\theta(\mathbf{x})$ is the local estimator for operator $\hat{O}$.
Since typical operators are $k$-local, the sum in the local estimator has a polynomial number of terms and can thus be efficiently computed.
An approximation to the ground (or low-energy) state of a system with Hamiltonian operator $\hat{H}$ is then found by minimizing the expectation value of the variational energy $E_\theta = \langle\hat{H}\rangle$ via gradient descent methods, such as stochastic reconfiguration~\cite{sorellaGeneralizedLanczosAlgorithm2001} or Adam~\cite{kingmaAdamMethodStochastic2017}, which we describe in more detail in Appendix \ref{app:optimization}.
The success (or otherwise) of VMC is thus related to the choice of three key components: 1) an expressive and compact ansatz; 2) a reliable sampling method and 3) a fast and robust optimization of the parameters.

Focusing on the first point, it can be important to impose physically motivated constraints on the chosen state in order to obtain a compact representation of the wave function, since one is typically interested in wave functions enforcing particular physical properties (e.g. those with an area scaling law in the entanglement entropy, topological order or antisymmetry for fermionic models).
However, in order to accurately model large extended systems, it is also critical that wave functions should be \textit{size extensive}.
This property requires that the error per particle incurred by the model in the asymptotic large system limit should remain constant with system size, ensuring that extensive thermodynamic quantities such as the energy density converge to a constant energy per unit volume.



\begin{figure}
    \includegraphics[width=\columnwidth,left]{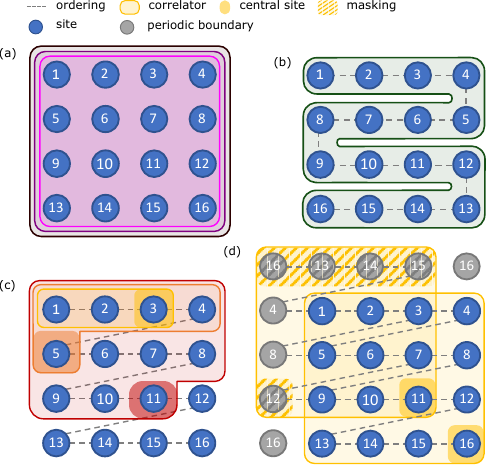}
    \caption{
        Schematic representations of the correlators used in different states on a 4$\times$4 lattice, showing the range of the explicit correlations in each:
        (a) a GPS with support dimension $M=3$ is a state with three correlators \toadd{each of which is parameterized on the occupation of sites over the whole system without restriction};
        (b) an MPS is obtained by introducing a one-dimensional ordering of the sites with direct entanglement approximated through sites in this order;
        (c) masking the correlators up to each site along a one-dimensional ordering and normalizing them leads to an autoregressive state, where each correlator is independently parameterized and depends only on the configuration of sites preceding it in the ordering (for clarity, only the correlators for sites $i=3,5,11$ are shown);
        (d) if the system has translational symmetry, the same correlator can be translated across the system, reducing the number of independent parameters in the model. This can be used in conjunction with the autoregressive model, as long as the site ordering is preserved by masking out `higher-ordered' sites in the filter as shown, to obtain an autoregressive filter-based state described in Sec.~\ref{subsec:filters} (for clarity, only the correlators for sites $i=11,16$ are shown). The range of these filters can be restricted if desired (not shown) to impose a further locality constraint.
    }
    \label{fig:ar-ansatze}
\end{figure}

A general guiding principle for size extensive parameterized wave functions is that they can be written in a product separable form as
\begin{equation}\label{eq:prod-sep-ansatz}
    \psi_{PS}(\mathbf{x}) = \prod_{i=1}^{N}\psi_i(\mathbf{x}),
\end{equation}
where $\psi_i(\mathbf{x})$ are individual parameterized functions (correlators) describing the $i$-th site and its correlations with other sites in its environment.
How these correlators are modelled has important consequences for the ability of the ansatz to capture different physical aspects of a wave function, such as the length scale or rank of the correlations it can model.
Simple product states have entirely local functions for each correlator, which precludes the description of multi-site correlated physics.
Recent ML-inspired variational ansätze such as the RBM or GPS have extensive correlators as long as the number of hidden units or support dimension respectively scales linearly with the system size, or if translational symmetries are taken into account.
Furthermore, full coupling of each site to their latent spaces of the model (`hidden layers' or `support states' respectively) allows each site to interact with the whole rest of the system, not formally restricting the rank of range of correlations that each site can describe, as illustrated in Figure \ref{fig:ar-ansatze}(a).
This enables them to capture entanglement scaling beyond the area law and to obtain accurate results formally independent of the dimensionality of the system~\cite{sunEntanglementFeaturesRandom2022,dengQuantumEntanglementNeural2017,sharirDeepAutoregressiveModels2020}.

In contrast, Matrix Product States (MPS) introduce a specific one-dimensional ordering of the degrees of freedom in a system, as shown schematically in Figure \ref{fig:ar-ansatze}(b), explicitly allowing for the efficient extraction of correlations decaying over finite length scales along the one-dimensional ordering~\cite{borinApproximatingPowerMachinelearning2020}.
Formalized through entanglement scaling arguments~\cite{eisertAreaLawsEntanglement2010a}, this makes MPS particularly suited for many one-dimensional systems.
More general families of tensor decomposed or factorized forms also exist which can unify these ansätze under the same mathematical framework~\cite{clarkUnifyingNeuralnetworkQuantum2018}. 
As will be seen in the next section, autoregressive quantum states also rely on the general product structure of Eq.~\ref{eq:prod-sep-ansatz}, but introduce ordering constraints in the correlator functions, with important ramifications for the expressivity of the ansatz.

\subsection{Universal construction of autoregressive quantum states}\label{subsec:uni-arqs}

Modelling the probability distribution of $N$ random variables $\mathbf{x}=(x_1,\dots,x_N)$ is a task common to many domains of science and engineering.
Advances in generative machine learning have popularized efficient approaches to describe and subsequently sample from a joint probability distribution $p(\mathbf{x})$ via an autoregressive (AR) factorization~\cite{bond-taylorDeepGenerativeModelling2022}.
This relies on the probability chain rule to decompose $p(\mathbf{x})$ into a product of conditional distributions $p(x_i|\mathbf{x}_{<i})$:
\begin{align}\label{eq:ar-fact}
    p(\mathbf{x}) &= \prod_{i=1}^N p(x_i|\mathbf{x}_{<i}) \\
    &= p(x_1)p(x_2|x_1)\cdots p(x_N|x_{N-1},\dots,x_2,x_1),
\end{align}
where $\mathbf{x}_{<i}=(x_1,\dots,x_{i-1})$ is a fixed, ordered sub-sequence of the random variables up to the $i$-th position.

The same autoregressive factorization can be applied to the wave function, where importantly, this form also naturally has the desired product separable structure shown in Eq.~\ref{eq:prod-sep-ansatz}.
We can define an autoregressive wave function as
\begin{align}
    \psi_{\theta}(\mathbf{x}) = \prod_{i=1}^N \psi_{i}(x_i|\mathbf{x}_{<i}), \label{eq:ar-real}
\end{align}
where $\psi_{i}(x_i|\mathbf{x}_{<i})$ is the conditional wave function over the $D$ local quantum states of the $i$-th site of the system, conditioned on $\mathbf{x}_{<i}$, a configuration of the sub-Hilbert space of all the sites before the $i$-th one in a one-dimensional ordering of the system.
Here $\theta$ denotes the (potentially complex) parameters of the autoregressive wave function, and the number of local quantum states will be $D=2$ for a spin-$\frac{1}{2}$ system, or $D=4$ for a fermionic system.

We desire a square-normalized autoregressive state, $\sum_{\bf{x}}|\psi_{\theta}({\bf x})|^2=1$, which can be achieved if all conditional wave functions are normalized for every possible sub-configuration $\mathbf{x}_{<i}$, i.e. $\sum_{x'=0}^{D-1}|\psi_i(x'|\mathbf{x}_{<i})|^2 = 1$.
This condition can be explicitly imposed on the state by normalizing the conditionals at the point of evaluation~\cite{sharirDeepAutoregressiveModels2020}.
This local normalization scheme is efficiently computable, since it only involves summing over local Fock states of the $i$-th site, with the global normalization for a given configuration just the product of the local normalization of the conditional states.
We can therefore always consider unnormalized models for the conditional wave functions $\tilde{\psi}_i(x_i|{\bf x}_{<i})$, and apply the normalization as the model is evaluated for a given ${\mathbf{x}}$. 
We note that this autoregressive property remains when the state is multiplied by any function $e^{i \phi({\bf{x}})}$ that models a complex phase. 



Thus, we can summarize the construction of an autoregressive ansatz into the following general recipe:
\begin{enumerate}
    \item define a one-dimensional ordering of the system, which is equivalent to picking unique site indices;
    \item choose a model for each unnormalized conditional wave function $\tilde{\psi}_i(x_i|\mathbf{x}_{<i})$;
    \item in the evaluation of the wave function for a given configuration, compute each conditional and their respective configuration-dependent normalization in the chosen ordering of the sites.
\end{enumerate}
This gives the general form for an autoregressive state as
\begin{equation}\label{eq:ar-ansatz-main}
    \psi_{AR}(\mathbf{x}) = \prod_{i=1}^N\frac{\tilde{\psi}_i(x_{i}|\mathbf{x}_{<i})}{\sqrt{\sum\limits_{x'=0}\limits^{D-1}|\tilde{\psi}_i(x'|\mathbf{x}_{<i})|^2}} ,
\end{equation}
which we depict schematically in Fig. \ref{fig:ar-ansatze}(c), with the conditional wave functions for a few sites shown to be conditioned only on the occupations of the sites preceding them in the chosen site ordering.
The property of the ansatz being systematically improvable to exactness holds as long as the models for each conditional are themselves universal approximators.
Recently introduced autoregressive ansätze have parameterized these conditional wave functions with machine learning-inspired models such as deep convolutional neural networks~\cite{sharirDeepAutoregressiveModels2020}, recurrent neural networks~\cite{hibat-allahRecurrentNeuralNetwork2020}, tranformers~\cite{zhangTransformerQuantumState2023}, or hybrid models that incorporate tensor networks with deep learning architectures~\cite{wuTensornetworkQuantumStates2023,chenANTNBridgingAutoregressive2023}.

We will consider a simpler construction, motivated from Bayesian kernel models rather than neural networks, using the recently introduced Gaussian process state (GPS) for each conditional~\cite{glielmoGaussianProcessStates2020,rathBayesianInferenceFramework2020,rathQuantumGaussianProcess2022,rathFrameworkEfficientInitio2023}.
Similar to neural network parameterizations, this model is a systematically improvable universal approximator for these conditionals, written in a compact functional form as  
\begin{equation}\label{eq:gps-ansatz}
    \psi_{GPS}(\mathbf{x}) = \exp{\left(\sum_{m=1}^M\prod_{i=1}^{N}\epsilon_{x_i,m,i}\right)},
\end{equation}
where $\epsilon$ is a tensor of adjustable parameters with dimensions $(D,M,L)$, with $M$ denoting the `support dimension', the single model hyperparameter that controls the expressivity of the ansatz.
Crucially, increasing $M$ enlarges the class of states that the GPS wave function can span systematically towards exactness, since it formally defines a set of product states on which a kernel model can be trained to support the description \cite{rathQuantumGaussianProcess2022}.
The exponential form of Eq.~\ref{eq:gps-ansatz} ensures product separability, and allows the model to capture entanglement beyond area law states.
This form can then be related to an infinite series of products of sums of unentangled states, as well as constructively recast into a deep feed-forward neural network architecture~\cite{rathQuantumGaussianProcess2022}.

We can use this GPS model as a parametric form for each conditional rather than the full state, to adapt the state definition to an autoregressive GPS ansatz (AR-GPS) as
\begin{equation}\label{eq:argps-ansatz}
    \psi_{AR-GPS}(\mathbf{x}) = \prod_{i=1}^N\frac{\tilde{\psi}_{i,GPS}(x_{i}|\mathbf{x}_{<i})}{\sqrt{\sum\limits_{x'=0}\limits^{D-1}\left|\tilde{\psi}_{i,GPS}(x'|\mathbf{x}_{<i})\right|^2}},
\end{equation}
where the autoregressive masking of the configuration $\bf{x}$ is explicitly enforced in the argument of each exponential by only multiplying parameters related to sites with index $j\leq i$. 
This full AR-GPS ansatz has $DMN(N+1)/2$ parameters since it is a product of normalized GPS models for each site, each with support dimension $M$ over successively larger Hilbert spaces.
Tempering this additional scaling with number of parameters compared to the parent GPS model will be considered via filters in Sec.~\ref{subsec:filters}.

To conclude this section we return to the general formulation of AR models, to stress that there are two conditions which must be enforced, both of which constrain the flexibility of the state compared to the parent parameterization.
These are:
\begin{enumerate}
    \item A specific site (orbital) ordering is imposed, with the conditional wave function for site $i$ only allowed to depend on $\bf{x}_<i$, i.e. the occupation of sites preceding it in this ordering.
    In the rest of this work, we will denote this step as a {\em masking} operation.
    Explicit dependence of the conditional from occupation changes of sites higher in this ordering are therefore excluded.
    It can thus also be expected that the flexibility of the state will be dependent on this ordering, as it is also commonly observed for tensor network representations relying on an enforced one-dimensional sequence of sites~\cite{cataldiHilbertCurveVs2021,luoAutoregressiveNeuralNetwork2022}.
    \item Each conditional is explicitly normalized over the $D$ local Fock states in the evaluation of any configurational amplitude.
    This constraint also reduces the expressivity of each conditional, and the overall state.
\end{enumerate}
This loss of flexibility is offset by the practical advantages of direct sampling of configurations from the state in statistical estimators.
The autoregressive masking and the explicit normalization in fact allow the generation of independent and identically distributed configurations directly from the underlying Born distribution $|\psi_{\theta}(\mathbf{x})|^2$, avoiding autocorrelation in path-dependent Markov chain construction via e.g. Metropolis sampling and potential loss of ergodicity.
\toadd{
    Furthermore, there are other applications beyond the variational optimization of ground states where the assurance of a normalized model is often advantageous. For instance, in quantum state tomography, normalized autoregressive models allow for the efficient optimization of negative-log-likelihood loss functions over data samples, while in real-time dynamics autoregressive models have also been powerful, though can require care in the choice of integration scheme~\cite{carrasquillaReconstructingQuantumStates2019,luoAutoregressiveNeuralNetwork2022,sinibaldiUnbiasingTimedependentVariational2023}.
}
Nevertheless, it is important to understand the loss of variational flexibility for AR models due to these two constraints, as well as the computational overheads compared to their parent parameterizations and increases in parameter number, to appropriately understand the trade-offs and whether this loss of flexibility can simply be compensated by a more complex model for the conditionals.
We will numerically investigate these questions and quantify the impact of the individual constraints in Sec.~\ref{sec:results_heisenberg} by comparing to the original (non-autoregressive) GPS model.




\subsection{Filters}\label{subsec:filters}

The general autoregressive ansatz in Eq.~\ref{eq:ar-ansatz-main}, as well as the specific AR-GPS model of Eq.~\ref{eq:argps-ansatz}, allows each site conditional correlator to be modelled independently.
While this increases the variational flexibility of the ansatz, it implies that the number of parameters scales as $\mathcal{O}(N^2)$.
For large systems, this can be prohibitively expensive, thus schemes that bring the scaling down become necessary.
We consider a scheme analogous to the approach of translationally invariant convolutional layers in neural network parameterizations~\cite{chooTwodimensionalFrustratedText2019}, which define local filters of correlation features and can be applied independently, or in conjunction with an autoregressive model, akin to how the PixelCNN model was used as an autoregressive quantum state in Ref.~\onlinecite{sharirDeepAutoregressiveModels2020}. 

If the system being studied has translational symmetry, then it is reasonable to model each conditional correlator centered at a given site as the same, with its dependence being on the distance from the current conditional site to the ones in its environment.
We can consider these conditional correlators then as filters that are translated to each site, describing the translationally equivalent correlations of that site with its environment.
This ensures that the quantum fluctuations between sites are the same across the system, and only depend on the relative distance between sites.
Furthermore, these filters can also be combined with an autoregressive state, as long as the autoregressive masking is applied on top to ensure that only the occupation of preceding sites is accounted for in each correlator.
We refer to this kind of autoregressive ansatz as the \textit{filter-based} model.
It should be stressed however that while the application of filters on unnormalized GPS states (which we term the `filter-GPS' model) trivially conserves translational symmetry, the masking operation on top of the filters will break this rigorous translational symmetry.


To consider the specific construction for an autoregressive filter-based GPS (`AR-filter-GPS'), we model the unnormalized conditional correlators as
\begin{equation} \label{eq:filter-ar-gps}
    \tilde{\psi}_i(x_i|\mathbf{x}_{<i}) = \exp{\left(\sum_{m=1}^M\prod_{\{\mathbf{r}\}} g_i(\sigma(\mathbf{r}),  \epsilon_{x_{\sigma(\mathbf{r})}, m, \mathbf{r}_i-\mathbf{r}})\right)}.
\end{equation}
To ensure that the form generalizes for lattices of different dimensions, we define the product over sites above not by their index, but rather via the set of $N$ vectors $\{\mathbf{r}\}$ to each site in the system.
The function $\sigma(\mathbf{r})$ then defines the mapping between the vector to the site, and the index of the site.
The tensor of variational parameters then depends on the occupation of the site at the position given by the vector ($x_{\sigma(\mathbf{r})}$), the support index ($m$) defining the latent space, and the relative distance between the central site of the conditional correlator and the site defined by the vector ($\mathbf{r}_i-\mathbf{r}$).
Note that this relative distance is the shortest distance taking into account the periodic boundary conditions.
The $g_i(j,x)$ function then controls the masking operation for the conditional of site $i$, required for the autoregressive properties, given by
\begin{equation}
    g_i(j,x)=\begin{cases}
    1, & \text{if $j>i$}.\\
    x, & \text{if $j\leq i$},
  \end{cases}
\end{equation}
thus masking out contributions from sites which have an index higher than the central site in the one-dimensional ordering.
This state is depicted schematically in Fig.~\ref{fig:ar-ansatze}(d) for a filter size which extends to the whole lattice size.


The consequence of the parameters being defined by relative site distances is that the total number of parameters is reduced by a factor $\mathcal{O}(N)$, yielding the same scaling as the parent (non-autoregressive) GPS model, at the expense that each site conditional is no longer independently parameterized.
A further reduction in number of parameters can then be simply achieved by introducing a range cutoff in the convolutional filters, i.e. setting a maximum distance in the range $|\mathbf{r}_i-\mathbf{r}|$ in Eq.~\ref{eq:filter-ar-gps}.
Practically, this restricts the range of the correlations that are modelled, and it is common to define range-restricted filters in e.g. CNN-inspired NQS studies \cite{chooTwodimensionalFrustratedText2019}.
The number of parameters in the state is then independent of system size for a fixed range of correlations.
However, in this work, we consider filters which extend over the whole lattice, and therefore do not restrict the range of the filters in each conditional to a local set of sites.

An alternative and simpler strategy to reduce the number of parameters in autoregressive models compared to the filter-based approach is simply to share the parameters between different conditional models, i.e. to remove the dependence on $i$ in the $\epsilon_{x_j,m,j}^{(i)}$ factor in the full AR-GPS model of Eq.~\ref{eq:argps-ansatz}.
This `weight-sharing' scheme has been considered in autoregressive models based on feed-forward neural networks, such as NADE~\cite{uriaNeuralAutoregressiveDistribution2016}, which inspired the ARQS in Ref.~\cite{sharirDeepAutoregressiveModels2020}.
While this weight-sharing scheme reduces the computational cost for sample generation and amplitude evaluation, it introduces a highly non-trivial relationship between subsequent correlators in the autoregressive sequence which can not directly be linked to physical intuition, making it hard to justify as a parameterization.
As a concrete example, in Appendix~\ref{app:ar-product-states}, we show a constructive demonstration of how the full AR-GPS with $M=1$ can exactly describe any product state.
However, the weight-sharing adaptation is generally expected to require $M=N$ to describe an arbitrary product state, representing a significant increase in the model complexity, even for the description of entirely unentangled states.
We therefore will not consider these weight-sharing AR-GPS models further.

A further technique to compress the full autoregressive approach is to exploit recurrent neural network-based AR models~\cite{hibat-allahRecurrentNeuralNetwork2020}.
These consist of a parameterized function (recurrent cell) that recursively compresses the environmental part of the physical configuration for each conditional, retaining the autoregressive character of the state.
Information about the previous sites is encoded in a hidden state vector, which is updated by the recurrent cell at each site.
From a modelling perspective, this approach is similar to a range-restricted AR-filter-GPS, but has the advantage of being able to learn a system-dependent description of this filter, instead of specifying it into the model {\em a priori}.
We will explore connections between these two approaches in future work.

Beyond the number of parameters, the computational cost of both generating statistically independent configurations from the AR-GPS wave function (according to $|\Psi(\mathbf{x})|^2$), and evaluating their amplitude, is given as $\mathcal{O}(N^2)$, since $N$ correlators of complexity $\mathcal{O}(N)$ must be computed. 
In AR-filter-GPS models with a range cutoff this reduces to $\mathcal{O}(NK^d)$, where $K$ is a measure of the linear length scales included, and $d$ denotes the dimensionality of the filters.
However, the dominant cost in a VMC calculation is often in the evaluation of the local energy, particularly in {\em ab initio} systems where the Hamiltonian in second quantization has in general a quartic number of non-zero terms (although locality arguments can reduce this to quadratic~\cite{rathFrameworkEfficientInitio2023}), and the model amplitudes must then be evaluated at all these connected configurations.
Here, it is possible to reduce the computational cost of evaluating the AR-GPS model at each of these configurations, by exploiting the fact that these configurations only differ by a small occupation change from a reference configuration.
This `fast updating' scheme for the AR-GPS, described in Appendix~\ref{app:fastupdate}, yields a reduction in the naive cost of computing the local energy by a factor of $\mathcal{O}(N)$ and is used in all results.



\subsection{Universality} \label{subsec:universality}
Given the functional forms introduced in the previous sections, it is important to consider to what degree they are able to exactly represent arbitrary quantum states, and thus be considered universal ansätze.
For the parent GPS (Eq. \ref{eq:gps-ansatz}), this property has been demonstrated in Ref. \cite{rathQuantumGaussianProcess2022}, and since the AR-GPS is a product of $N$ GPS models this ansatz is also universal, even in the presence of the masking and normalization.

For the filter-GPS where translationally-invariant filters are used on top of the GPS, the ansatz will only be able to represent quantum states with trivial translational symmetry with character one, i.e. those where all translations map configurations onto ones with the same amplitude.
As such, the filter-GPS can only be considered a universal ansatz for states exhibiting this trivial translational symmetry, and as long as no locality constraints are applied to the filter, which is allowed to span the whole system (or deep architectures used as in Ref.~\cite{chooFermionicNeuralnetworkStates2020}). 
However, applying the autoregressive adaptation on top of a filter-based ansatz allows the symmetry to be broken by the masking operation, and while this no longer exactly conserves translational symmetry, it also allows the state to become a general universal approximator.
Masking can also be applied to filter-GPS for systems without exact translational symmetry, to break the enforcing of this symmetry by the filters and return to a universal approximator for all states (e.g. for lattice models with open boundary conditions)~\cite{zhouUniversalityDeepConvolutional2020}.

\subsection{Symmetries and conserved quantities} \label{subsec:symmetrization}

Incorporating symmetries and conserved quantities of the system into the ansatz is crucial for state-of-the-art accuracy~\cite{nomuraHelpingRestrictedBoltzmann2021,chooTwodimensionalFrustratedText2019}, restricting the optimization to the appropriate symmetry sector.
Any non-autoregressive GPS state can typically be symmetrized by either symmetrizing the form of the kernel via a filter as described in Sec.~\ref{subsec:filters} (and as has previously been denoted `kernel symmetrization as in Eq. (B1) of Ref.~\onlinecite{rathQuantumGaussianProcess2022}), or via projective symmetrization where an operator summing over the operations of the group is applied at the point of evaluating the amplitudes (see Eq.~(B2) in Ref.~\onlinecite{rathQuantumGaussianProcess2022}).
For a set $\mathcal{S}$ of symmetry operations forming a symmetry group, the projective symmetrization sums over the symmetry operations applied to each configuration that is being evaluated of a non-symmetric ansatz, $\psi_{\theta}({\mathbf x})$.
Projecting onto a specific irreducible representation of the group, $\Gamma$, results in the explicitly symmetrized ansatz
\begin{equation}
    \psi_{\theta}^{\Gamma}(\mathbf{x}) = \frac{{\text{dim}}(\Gamma)}{|\mathcal{S}|}\sum_{\tau\in\mathcal{S}} \chi^{\Gamma}(\tau) \psi_{\theta}(\tau(\mathbf{x})), \label{eq:proj-sym}
\end{equation}
where each $\tau$ is a symmetry operation, and $\chi^{\Gamma}(\tau)$ is the character of the symmetry operation in that irrep.
For the totally symmetric states considered in this work all characters are one, leading to a simple averaging over the symmetry-related configurations~\cite{nomuraHelpingRestrictedBoltzmann2021,rothGroupConvolutionalNeural2021, nomuraDiracTypeNodalSpin2021,rothHighaccuracyVariationalMonte2023}. 

However, since this explicit symmetrization of Eq.~\ref{eq:proj-sym} does not preserve the normalization of the state, projective symmetrization is not compatible with the autoregressive property, and thus direct sampling of configurations.
Instead, autoregressive wave functions are symmetrized by ensuring that the probability of generating configurations of a certain symmetry-equivalence class from the unsymmetrized ansatz is the same as the probability given by the corresponding amplitude of the symmetrized model.
As first proposed for autoregressive NQS in Ref.~\onlinecite{sharirDeepAutoregressiveModels2020} and further expanded on in Ref.~\onlinecite{rehOptimizingDesignChoices2023}, this can be achieved by averaging the real and imaginary part of the wave function amplitude separately, which for a totally symmetric irrep results in
\begin{multline}\label{eq:symm-ar-ansatz}
    \psi_{\theta}^{\text{symm}}(\mathbf{x}) = \sqrt{\frac{1}{|\mathcal{S}|}\sum_{\tau\in\mathcal{S}}\exp{\left(2\Re{\left[\log\psi_{\theta}(\tau(\mathbf{x}))\right]}\right)}} \\
    \times\exp{\left(i\arg{\left(\sum_{\tau\in\mathcal{S}}\exp{\left(i\Im{\left[\log\psi_{\theta}(\tau(\mathbf{x}))\right]}\right)}\right)}\right)},
\end{multline}
where $\psi_{\theta}(\mathbf{x})$ is the amplitude from the unsymmetrized autoregressive ansatz.
This ansatz can be sampled in a two-step process: first a configuration is autoregressively sampled from the unsymmetrized ansatz, then a symmetry operation is drawn uniformly from the set $\mathcal{S}$ and applied to the sampled configuration.

For symmetry operators which are diagonal in the computational basis, there is a simpler way to exactly constrain the sampled irrep in autoregressive models.
This includes spin magnetization when working in a computational basis of ${\hat S}^z$ eigenfunctions, or electron number symmetry for fermionic models.
This can also be extended to full $\mathrm{SU}(2)$ spin-rotation symmetry when working in a basis of coupled angular momentum functions~\cite{vieijraRestrictedBoltzmannMachines2020,luoGaugeinvariantAnyonicsymmetricAutoregressive2023}.
These `gauge-invariant' autoregressive models can be implemented with a gauge-checking block, which renormalizes the conditionals in order to respect the overall selected gauge or quantum number.
This means that certain local Fock states of conditionals are set to zero when iteratively generating a configuration, if they would result in a symmetry-breaking final configuration.
Therefore, this excludes support of the AR-GPS state on these symmetry-breaking configurations.
This approach has also been used for autoregressive recurrent neural network-based architectures for the conservation of magnetization in quantum spin systems~\cite{hibat-allahRecurrentNeuralNetwork2020} and electron number and multiplicity in {\em ab initio} systems~\cite{barrettAutoregressiveNeuralnetworkWavefunctions2022,malyshevAutoregressiveNeuralQuantum2023}.



In this work, we use the normalization-preserving symmetrization of Eq.~\ref{eq:symm-ar-ansatz} to symmetrize our AR-GPS with the $C_{4v}$ point-group symmetries of the lattice and $\mathbb{Z}_2$ spin-flip symmetry in quantum spin systems (not including translations, which are not preserved even in the presence of filters in the AR states). 
In addition, we implement a gauge-checking block to conserve the total magnetization in spin systems as well as the electron particle number in fermionic systems.
We stress here that the normalization-preserving symmetrization of Eq.~\ref{eq:symm-ar-ansatz} is not equivalent to the projective symmetrization approach of Eq.~\ref{eq:proj-sym}.
In particular, in keeping with the conclusions of Refs.~\onlinecite{rehOptimizingDesignChoices2023} and ~\onlinecite{rothHighaccuracyVariationalMonte2023}, we find that the symmetric autoregressive state resulting from Eq.~\ref{eq:symm-ar-ansatz} is not as capable in modelling sign structures of quantum states compared to the projective symmetrization of non-autoregressive states.
This is due to the requirement to split the amplitude and phase information in Eq.~\ref{eq:symm-ar-ansatz}, which prevents interference between unsymmetrized amplitudes, limiting the flexibility of autoregressive ansätze in frustrated spin and fermionic systems.


\section{Results}\label{sec:results}

Having presented the general formulation of autoregressive and filter adaptations to a wave function ansatz, as well as their specific construction for the Gaussian process state (GPS) model, we will now numerically investigate the expressivity of these states.
In particular, we aim to understand how the AR constraints of masking conditionals according to a 1D ordering and normalization (Sec.~\ref{subsec:uni-arqs}), as well as the symmetrization (Sec.~\ref{subsec:symmetrization}) and convolutional filters (Sec.~\ref{subsec:filters}) change the variational freedom of the state compared to the `parent' unnormalized and non-autoregressive GPS model of Eq.~\ref{eq:gps-ansatz}.
Note that this is analyzed independently to the benefit in the efficiency of the direct sampling afforded by the AR models, which is considered elsewhere~\cite{sharirDeepAutoregressiveModels2020} and likely to be highly system dependent.
We therefore consider the minimized variational energy of these models, with the complexity of the state denoted by the number of parameters.
These models are optimized with a variant of the stochastic reconfiguration algorithm~\cite{sorellaGeneralizedLanczosAlgorithm2001,lovatoHiddennucleonsNeuralnetworkQuantum2022} which is detailed further in Appendix~\ref{app:optimization} and has been shown to improve the convergence for autoregressive models and avoid local minima.
We implement all the models of Table \ref{tab:ansatze} and perform the VMC optimization using the NetKet package~\cite{carleoNetKetMachineLearning2019,vicentiniNetKetMachineLearning2022}.

To distinguish between the effects of the autoregressive masking and the normalization conditions, we also consider one further GPS-derived model.
This is an unnormalized ansatz, but including the autoregressive masking, as
\begin{equation}\label{eq:masked-ansatz}
    \psi_{\textrm{masked-GPS}}(\mathbf{x}) = \prod_{i=1}^N\exp{\left(\sum_{m=1}^M\prod_{j\leq i}\epsilon_{x_j,m,j}^i\right)}.
\end{equation}
This `masked-GPS' model is purely proposed for illustrative purposes, as it suffers from the increase in variational parameters and loss of flexibility due to masking constraints of the full AR-GPS state, but without the benefit of direct sampling that the full AR construction would afford.
However, since it does not introduce the additional normalization constraint, it allows us to disentangle the loss in the flexibility of the model due to these two constraints required for a full AR state construction.
We also apply this masking without the explicit normalization in the presence of filters, resulting in the `masked-filter-GPS' similarly used to understand the effect of the masking operation in isolation.

\begin{table}
\resizebox{\columnwidth}{!}{%
\begin{tabular}{l c c c c}
    \hline
    Ansatz & Masked & Normalized & Correlator & Parameters\\
    \hline
    GPS & \xmark & \xmark & fully-variational & $\mathcal{O}(DMN)^{\dagger}$ \\
    masked-GPS & \cmark & \xmark & fully-variational & $\mathcal{O}(DMN^2)$ \\
    AR-GPS & \cmark & \cmark & fully-variational & $\mathcal{O}(DMN^2)$ \\
    filter-GPS & \xmark & \xmark & filter & $\mathcal{O}(DMN)$ \\
    masked-filter-GPS & \cmark & \xmark & filter & $\mathcal{O}(DMN)$ \\
    AR-filter-GPS & \cmark & \cmark & filter & $\mathcal{O}(DMN)$ \\
\end{tabular}
}
\caption{
    The variational GPS variants used in this work for the comparison between autoregressive and non-autoregressive states and their respective properties, including how the number of parameters scales w.r.t. the dimension of the local Fock space $D$, the support dimension $M$ and the system size $N$.
    `Filter' correlators impose that all sites model identical correlations with their environment (up to masking constraints), while fully-variational correlators allow for fully-independent correlators for each site.
    Note that the masked-filter-GPS and masked-GPS are models used to understand and compare the effect of certain properties, but are not expected to be used in practice.
    $\dagger$: In order to achieve product-separability, the support dimension $M$ is expected to scale as $\mathcal{O}(N)$.
}
\label{tab:ansatze}
\end{table}

We first consider these states applied to the Heisenberg system, before moving on to fermionic Hubbard models, and {\em ab initio} systems, to understand the change of variational freedom that these model variants present.
While these will be specific to the GPS underlying parameterization, we expect that the conclusions regarding the relative expressibility of these states will also transfer to other (e.g. NQS) parent models with similar universal approximator properties.



\subsection{Antiferromagnetic Heisenberg system}
\label{sec:results_heisenberg}
The study of quantum spin fluctuations and magnetic order in condensed matter has for a long time relied on the understanding of the properties of the $S=1/2$ antiferromagnetic Heisenberg model (AFH), described by ${\hat H} = J\sum_{\langle i,j\rangle} \mathbf{{\hat S}}_i\cdot\mathbf{{\hat S}}_j$, where $\langle i,j\rangle$ represent nearest-neighbor pairs of localized quantum spins.
Correctly describing the quantum fluctuations of the AFH ground state at zero temperature represents a challenge for analytical and numerical techniques and a common benchmark for many emerging methods.
On a $6 \times 6$ square lattice,
we can apply the Marshall-Peierls sign rule, transforming the Hamiltonian into a sign-free problem that can be described with real parameter GPS models, avoiding in this case the complications of representing sign-structures as described in Sec.~\ref{subsec:symmetrization}.
For masked and autoregressive models, we follow a zig-zag ordering of the lattice sites, as depicted in Fig. \ref{fig:ar-ansatze}(c-d).

\begin{figure}
    \includegraphics[width=\columnwidth]{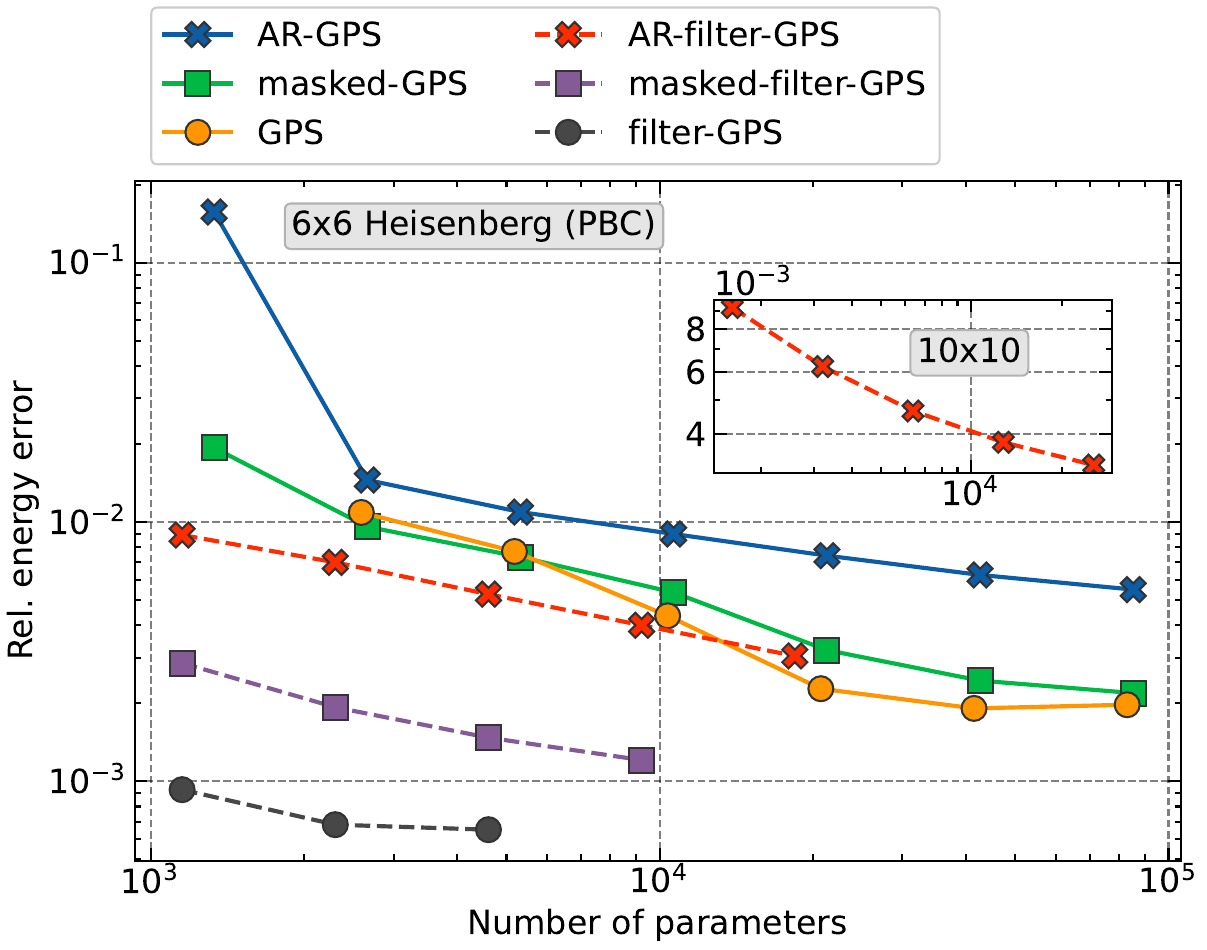}
    \caption{
        Relative energy error of different GPS models for the $6\times 6$ AFH with PBC as a function of the number of parameters.
        The exact energy is obtained from Ref.~\onlinecite{schulzMagneticOrderDisorder1996}.
        Models with convolutional filters are shown with dashed lines and are more parameter efficient, while models with independent site conditionals are shown with solid lines.
        We find that explicit normalization of the autoregressive state is more deleterious to its variational freedom than the other AR requirement of masking conditionals based on a 1D ordering of the lattice sites, for both filter and non-filter models.
        The explicitly translationally-invariant filter-GPS without masking or normalization constraints are found to be the variationally best ansatz for a given complexity in this system.
        Inset: Relative energy error of the AR-filter-GPS model for the larger $10 \times 10$ lattice compare to the stochastic series expansion results of Ref.~\onlinecite{sandvikFinitesizeScalingGroundstate1997}.}
    \label{fig:heisenberg2d-rel-error}
\end{figure}

In Fig. \ref{fig:heisenberg2d-rel-error} we show the relative variational energy error as a function of the number of parameters for the different ansätze.
We include the conservation of total zero magnetization for all models (trivially for the Metropolis-based non-AR models, or via gauge-checking blocks for the AR models), and do not explicitly include further symmetries unless otherwise stated.
The accuracy for all states can be systematically improved \toadd{in principle} by increasing the number of parameters (via the support dimension $M$), noting that due to the difference in scaling the same number of parameters does not necessarily equate to the same value of $M$. \toadd{We take care to ensure that the models are optimized as well as possible with respect to samples and other technical parameters (see Appendix A for more details), such that we can accurately judge the overall ground state expressibility of the models. However we can not exclude the possibility of the AR models changing the optimization landscape to introduce fundamental bottlenecks in the training as an alternative source of the discrepancies.}

Considering first the GPS models without filters (solid lines), there is a clear loss of variational flexibility for a given number of parameters between the unnormalized parent GPS model (Eq.~\ref{eq:gps-ansatz}), and the AR-GPS model (Eq.~\ref{eq:ar-ansatz-main}).
Interestingly, if we apply a masking operation to the GPS model without normalization (the non-AR masked-GPS model of Eq.~\ref{eq:masked-ansatz}), the energies are almost as good as the parent GPS model, despite the increase in the number of parameters for a given $M$.
This indicates that it is the act of explicitly normalizing the AR-GPS state for each configuration which is providing most of the loss of variational flexibility in the model, rather than the act of masking the physical configuration from certain sites.
This normalization step cannot be easily compensated for by an increase in the support dimension.
We note here that similar finding have been uncovered in the machine learning literature, pointing to intrinsic limitations of autoregressive models in modelling arbitrary distributions over sequences of a finite length~\cite{linLimitationsAutoregressiveModels2021,wangYourAutoregressiveGenerative2022}.

We further consider the relative impact of this masking and normalization of the individual conditionals of the autoregressive state, but now with filters also applied both to autoregressive, masked and parent GPS models (dashed lines).
While these models are more parameter efficient than their respective counterparts without filters, the discrepancy between the AR-filter (including normalization of conditionals) and the masked-filter-GPS (without normalization of conditionals) persists, reaffirming that the normalization rather than the masking is the leading cause in the loss of flexibility going towards AR models in this (unsigned) problem.
As expected, the filter-GPS provides the best results for a given number of parameters, due to its dual advantage in both avoiding the masking operation, allowing the model at each site to `see' the full spin configuration, as well as ensuring that translational symmetry is exactly maintained.
The quality of these results in this system mirrors the equivalent `kernel-symmetrization' results of Ref.~\onlinecite{rathQuantumGaussianProcess2022}.
We should stress again that this analysis considers purely the expressivity of these models for a given compactness, rather than the numerical advantages in faithful and direct sampling of configurations the AR construction admits.

In the inset of Fig.~\ref{fig:heisenberg2d-rel-error} we also report the relative energy error obtained by the filter-based autoregressive GPS model on a larger $10\times 10$ lattice.
This relative error is almost identical to that reached on the smaller $6\times 6$ lattice with the same value of $M$, confirming the expectation of a consistent level of accuracy across different system sizes for a given $M$ for the size extensive form of the AR-GPS model.

\begin{figure}
    \includegraphics[width=\columnwidth]{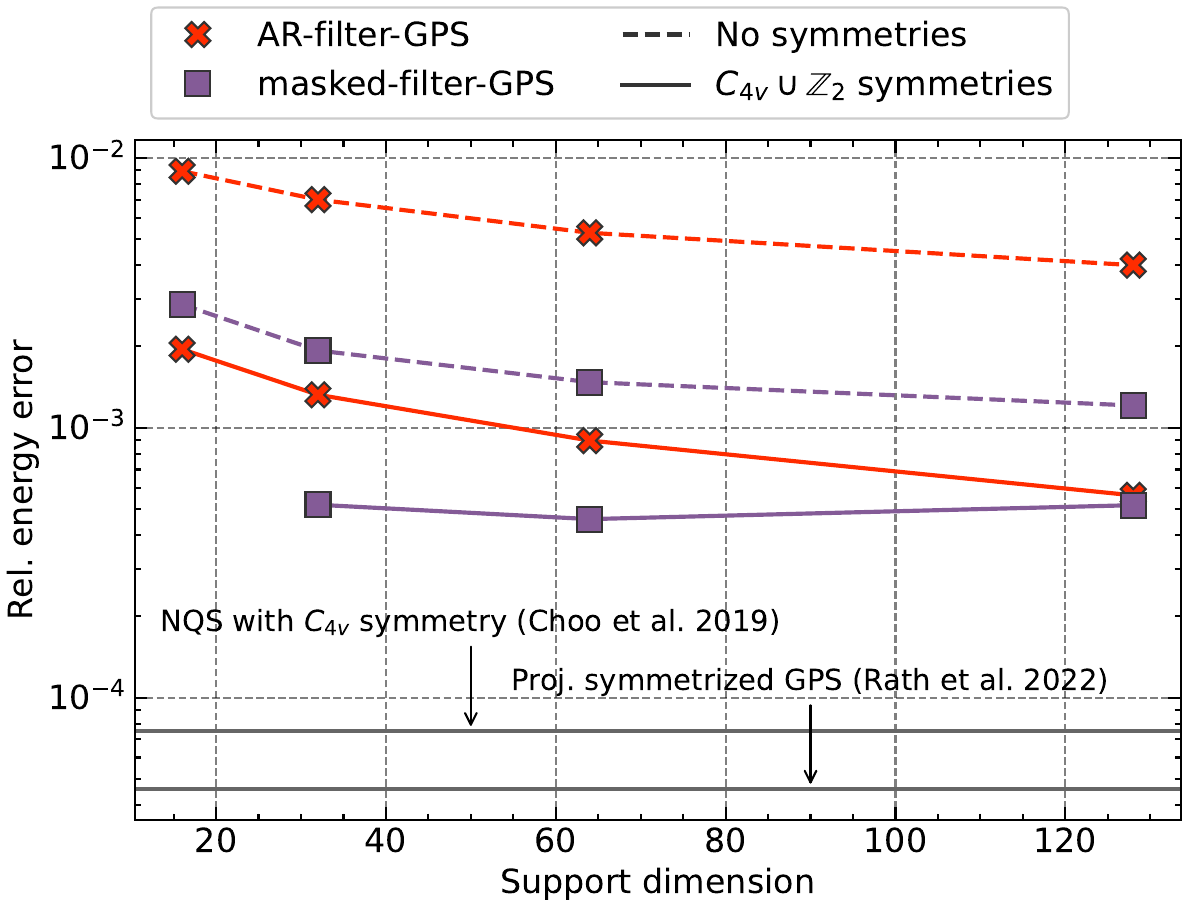}
    \caption{Relative energy error of the autoregressive (red) and masked (violet) filter-based GPS ansätze without symmetrization (dashed) and with normalization-preserving $C_{4v}$ and $\mathbb{Z}_2$ symmetrization (solid) as a function of the support dimension ($M$) on the $6\times 6$ AFH with PBC (conservation of total zero magnetization is included in all models).
    Included is also the relative energy error of a projectively-symmetrized (non-autoregressive) CNN-based NQS~\cite{chooTwodimensionalFrustratedText2019}, which provides similar accuracy and compactness to the projectively-symmetrized GPS \cite{rathQuantumGaussianProcess2022} model due to the improved symmetrization and lack of masking.
}
    \label{fig:heisenberg2d-symmetries}
\end{figure}

To test whether the inclusion of additional symmetries helps in closing this accuracy gap due to the explicit normalization of AR correlators, we optimize filter-based autoregressive and masked GPS models with the inclusion of $C_{4v}$ point-group symmetries of the square lattice and $\mathbb{Z}_2$ spin-flip symmetry (in addition to the conservation of total zero magnetization, i.e. $S_z$ symmetry).
We symmetrize both models following the normalization-preserving method in Eq.~\ref{eq:symm-ar-ansatz}, in order to ensure a faithful comparison.
Inclusion of these symmetries in Fig.~\ref{fig:heisenberg2d-symmetries} show that they help with the accuracy of both models.
However, the gap between the accuracy of the models (arising from the requirement of explicit normalization of conditional correlators in the AR-filter-GPS) decreases as a function of $M$.
This is due to a plateau in the accuracy of the masked-filter-GPS model.

For comparison, we also show the CNN-based multi-layer NQS results from Ref.~\onlinecite{chooTwodimensionalFrustratedText2019}.
While this non-AR filter-based NQS model is by construction translationally invariant, it is also invariant under all rotations of the lattice ($C_4$ symmetry).
However, contrary to the masked and autoregressive filter-based GPS models in Fig.~\ref{fig:heisenberg2d-symmetries}, the CNN-NQS was symmetrized via projective-symmetrization, which as shown in Ref.~\onlinecite{rehOptimizingDesignChoices2023} results in a more expressive model than the normalization-preserving symmetrization required in autoregressive models, which is again demonstrated in these results.
Furthermore, even though our models use a translationally-invariant filter to model the conditional wave-functions, the autoregressive masking breaks this invariance, and thus we lose this exact symmetry in the model.
Restoring rigorous translational symmetry in the AR-filter-GPS models would require the addition of all the translation operators in Eq.~\ref{eq:symm-ar-ansatz}, yielding an additional $\mathcal{O}(N)$ cost to the evaluation of the amplitudes \toadd{unless working within a basis representation which transforms according to the symmetry group \cite{malyshevAutoregressiveNeuralQuantum2023,barrettAutoregressiveNeuralnetworkWavefunctions2022}.}

We expect that this exact conservation of translational symmetry, more expressive (projective) symmetrization of the point group operations, as well as lack of masking to be the dominant cause of the improved CNN results of Ref.~\onlinecite{chooTwodimensionalFrustratedText2019} rather than the change in underlying model architecture to the GPS in Fig.~\ref{fig:heisenberg2d-symmetries}.
This is validated via inclusion of the projectively-symmetrized GPS results of Ref.~\onlinecite{rathQuantumGaussianProcess2022}, which provides comparable accuracy to the projectively-symmetrized deep CNN results of Ref.~\onlinecite{chooTwodimensionalFrustratedText2019} (albeit noting that the GPS results are also projectively-symmetrized over the translational symmetries, instead of relying on translationally-invariant filters). \toadd{We would also expect other more recent NQS architectures to be able to efficiently model this system \cite{viterittiTransformerVariationalWave2023}.}


\subsection{1D Hubbard model}\label{subsec:hubbard1d}
\begin{figure}
    \includegraphics[width=\columnwidth]{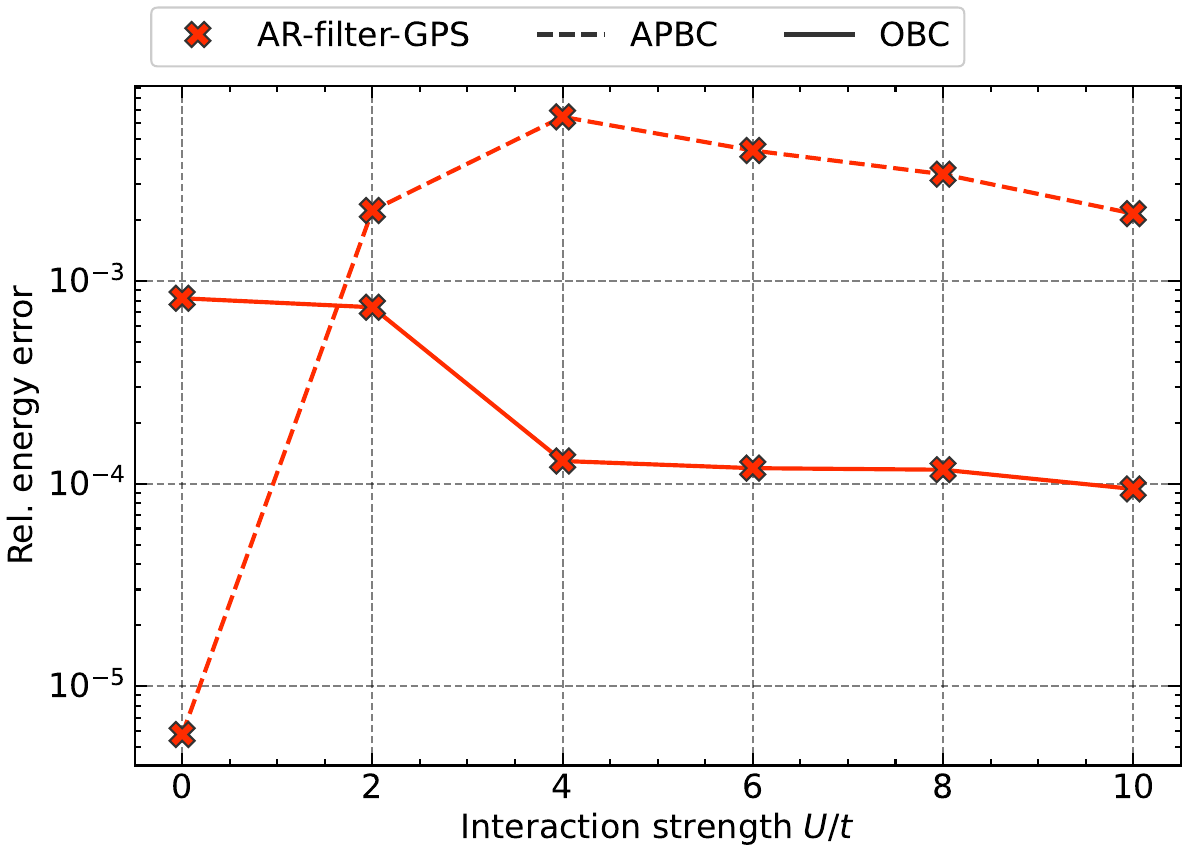}
    \caption{Relative energy error of the autoregressive filter-based GPS model with support dimension $M=64$ as a function of the interaction strength $U/t$ in a 1D Hubbard system with $N=32$ sites and different boundary conditions.
    Solid lines represent models optimized with Open Boundary Conditions (OBC), while dashed lines are for those obtained with Anti-Periodic Boundary Conditions (APBC).
    Exact reference energies were obtained from an MPS with bond dimension $M=2500$ optimized by DMRG~\cite{zhaiLowCommunicationHigh2021}.}
    \label{fig:hubbard1d-rel-error}
\end{figure}

We now move to the 1D fermionic Hubbard model of strongly correlated electrons with Hamiltonian
\begin{equation}
    H = -t\sum_{\langle i,j\rangle,\sigma}\left(\hat{c}^{\dagger}_{i\sigma}\hat{c}_{j\sigma}+\hat{c}^{\dagger}_{j\sigma}\hat{c}_{i\sigma}\right) + U\sum_i\hat{n}_{i\uparrow}\hat{n}_{i\downarrow},
\end{equation}
where $\hat{c}^{\dagger}_{i\sigma}(\hat{c}_{i\sigma})$ is the creation (annihilation) operator for a $\sigma$-spin electron at site $i$, and $\hat{n}_{i\sigma}=\hat{c}^{\dagger}_{i\sigma}\hat{c}_{i\sigma}$ is the spin-density operator for $\sigma$-spin electrons at site $i$~\cite{arovasHubbardModel2022}.
The ansätze introduced can be easily extended to this fermionic setting by allowing the local Fock space for each site to be extended to four possible states, from two in spin systems.
In Fig.~\ref{fig:hubbard1d-rel-error} we show the relative ground state energy error of a AR-filter-GPS with support dimension $M=64$ for a $N=32$ site 1D model in different interaction regimes, from the uncorrelated $U=0t$ to strongly correlated $U=10t$, compared to the reference energy from a DMRG optimized MPS with bond dimension $M=2500$~\cite{zhaiLowCommunicationHigh2021}. 

For each interaction strength we consider both open (OBC) and anti-periodic (APBC) boundary conditions.
The OBC system no longer strictly obeys translational symmetry, however the filter ensures that the environment around each site is modelled with the same parameters.
In this case, the sum over $\{\bf{r}\}$ in Eq.~\ref{eq:filter-ar-gps} therefore only ranges over values such that $\mathbf{r}_i-\mathbf{r}$ respects the boundary conditions of the system for each site conditional.
As discussed in Sec.~\ref{subsec:universality}, despite the filter ensuring that environmental fluctuations are translationally symmetric, the addition of the masking operation ensures that the AR-filter-GPS is still a universal approximator for this system even with OBC.

In fact, we find that the model is able to describe the OBC system in general to higher accuracy than the APBC system, with it being particularly effective at higher $U/t$ values.
We can rationalize this as resulting from the fact that the necessity of the masking operation biases towards the OBC, where a 1D ordering of the sites is required which does not respect the boundary conditions of the problem.
This improves at higher $U/t$ values, where the physics is dominated by local fluctuations, and where the imposition of a translationally symmetric filter is less restrictive.
In contrast, the translational symmetry of the APBC system works against the constraints imposed by the masking, where the accuracy is worse for all values other than the uncorrelated $U/t=0$ state.


\subsection{Ab initio hydrogen models}
Lastly we look at more challenging {\em ab initio} fermionic systems with long-range Coulomb interactions, and test the performance of the autoregressive GPS in describing the electronic ground state of chains and sheets of hydrogen atoms.
Analogous to changes in the interaction strength in the Hubbard system, modulating the bond length of the hydrogen atoms leads to qualitatively different correlation regimes.
These systems have been studied as models towards realistic bulk materials, with rich phase diagrams that exhibit Mott phases, charge ordering and insulator-to-metal transitions~\cite{rathFrameworkEfficientInitio2023,simonscollaborationonthemany-electronproblemSolutionManyElectronProblem2017,hachmannMultireferenceCorrelationLong2006,stellaStrongElectronicCorrelation2011,sinitskiyStrongCorrelationHydrogen2010,tsuchimochiStrongCorrelationsConstrainedpairing2009}.

In second quantization, the \textit{ab initio} Hamiltonian in the Born-Oppenheimer approximation is discretized in a basis of $L$ spin-orbitals
\begin{equation}
    \hat{H} = \sum_{ij}^{2L}h^{(1)}_{ij}\hat{c}^{\dagger}_i\hat{c}_j + \frac{1}{2}\sum_{ijkl}^{2L}h^{(2)}_{ijkl}\hat{c}^{\dagger}_i\hat{c}^{\dagger}_k\hat{c}_j\hat{c}_l,
    \label{eq:ab-initio_ham}
\end{equation}
where $\hat{c}^{\dagger}_i(\hat{c}_i)$ are fermionic operators that create (destroy) an electron in the $i$-th spin-orbital.
The single-particle contributions due to the kinetic energy of the electrons and their interaction with external potentials are described by the one-body integrals $h^{(1)}_{ij}$, whereas the Coulomb interactions between electrons is modelled by the two-body integrals $h^{(2)}_{ijkl}$.

The choice of molecular orbitals used in the second quantized representation is not unique, since any non-singular rotation of the orbitals would yield another valid basis for the degrees of freedom, without affecting the physical observables of the exact solution.
However, depending on the chosen representation for the computational basis, the wave function will have different amplitudes, which changes the ability to sample configurations from an ansatz, as well as faithfully represent them in a given parameterized form.
As such the accuracy to which an observable can be estimated by an ansatz will greatly depend on this choice, which will then also impact the optimization process.
The practical consequences of this choice on the scalability of VMC calculations have recently been studied in Ref.~\onlinecite{rathFrameworkEfficientInitio2023} with the GPS as an ansatz.
Using the $4\times 4\times 4$ hydrogen cube as benchmark, the authors have demonstrated the benefits of working in a basis of localized orbitals to obtain state-of-the-art results.

A localized basis is one in which the electron orbitals are rotated in such a way that they fulfil some locality requirement, concentrating the orbital amplitudes around localized regions in the system, often retaining atomic-like orbital character.
In contrast, the orbitals in a more common canonical basis (that diagonalize some single-particle effective Hamiltonian, such as the Hartree--Fock or Kohn--Sham Hamiltonian) are delocalized over the whole system.
Since the dominant contribution to the correlated ground state wave function typically would come from the mean-field configuration in this representation, the probability distribution over configurations is generally highly peaked around this configuration.
For a local basis, however, many configurations will have similar energy contributions, which then leads to more uniform structure in the probability distribution, improving the ability to faithfully sample from the wave function via Monte Carlo algorithms.

It is important to note here that even though autoregressive states allow for independent and uncorrelated sampling of the configurations, they are not immune to sampling effects caused by the choice of representation.
In a canonical basis, they will still require potentially many samples to resolve expectation values by integrating over the highly peaked probability distribution given by the ground state many-body density, which is the same problem that can also affect non-autoregressive states as described in Ref.~\onlinecite{chooFermionicNeuralnetworkStates2020}.
Autoregressive states are simply more sample efficient, since generated configurations  are uncorrelated, and amplitudes of repeatedly generated configurations can be stored in a lookup table as illustrated in the batched sampling procedure of Ref.~\cite{barrettAutoregressiveNeuralnetworkWavefunctions2022}.

\begin{figure}
    \includegraphics[width=\columnwidth]{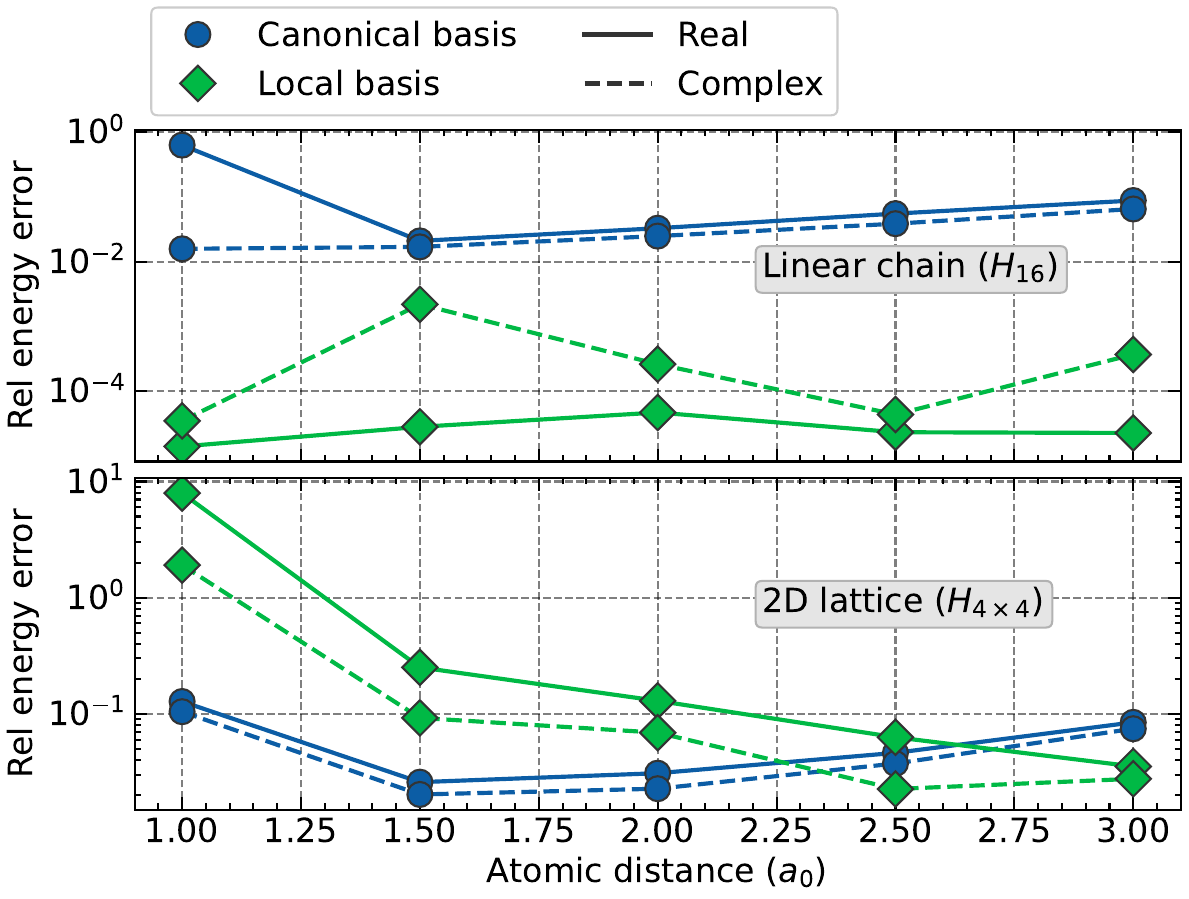}
    \caption{
        Relative energy error of an $M=16$ autoregressive GPS on a chain of 16 (top) and a $4\times 4$ lattice (bottom) of hydrogen atoms at different interatomic distances, using a canonical (blue circles) and a local (green diamonds) basis representation, in an underlying STO-6G basis.
        Solid and dashed lines represent states optimized with real and complex parameters respectively.
        Exact energies are obtained via full configuration interaction algorithm from \texttt{PySCF}~\cite{sunPySCFPythonbasedSimulations2018,sunRecentDevelopmentsPySCF2020}.
    }
    \label{fig:H16-rel-error}
\end{figure}

In Fig.~\ref{fig:H16-rel-error} we investigate this by considering the relative energy error of a fully-variational AR-GPS model with support dimension $M=16$ on 1D chains with OBC and $4 \times 4$ square planar 2D systems of 16 hydrogen atoms (32 spin-orbitals) in both local and canonical (restricted Hartree--Fock) bases at different interatomic distances~\cite{sunPySCFPythonbasedSimulations2018,sunRecentDevelopmentsPySCF2020}.
We then obtain a localized basis by performing a Foster-Boys localization~\cite{fosterCanonicalConfigurationalInteraction1960}, which directly minimizes the overall spatial extent of each orbital, whilst preserving orthogonality.

The 1D hydrogen chain can be considered an extension of the 1D Hubbard model with OBC of Sec.~\ref{subsec:hubbard1d}, with a natural choice of ordering for the local orbitals required for the masking, but now with long-range interactions giving the potential to induce a non-trivial sign structure of the state, and a higher complexity in the local energy evaluation.
As shown in the top panel of Fig. \ref{fig:H16-rel-error}, the localization of the orbitals in this setting is clearly critical for the sampling of configurations in the optimization of the autoregressive model, as well as the accuracy to which the state can be represented.
In the local basis, the ansatz achieves an average relative energy error of $\approx 4\times 10^{-5}$ across the whole range of interatomic distances considered, whereas in the canonical basis it fails to reach an acceptable accuracy, with its error increasing as the separation between atoms becomes large in the more strongly correlated regime.

We furthermore tested the performance of the AR-GPS ansatz with both real-valued parameters, restricting it to model positive-definite wave function amplitudes, as well as complex parameters, which should enhance the flexibility while also making it possible to model the sign structure of the target state.
The additional freedom of the model with complex parameters leads to an improvement in the observed energy in all cases other than the linear chain in a local basis.
In this case, the high accuracy of the real-valued, strictly positive model could not be matched.
Since the complex-parameter model must be able to span the same states as the real-valued analogue, this (small) discrepancy must arise from increased difficulties in the sampling and optimization of the parametrization, even if the model carries the theoretical ability to give a better approximation.
The additional flexibility of the complex amplitudes causes more numerical difficulties in practice than benefit found in their expressibility, due to the small change from a stoquastic Hamiltonian and positive-definite wave function that the long-range interactions induce in this case.

Rearranging the atoms into a two-dimensional square lattice changes these conclusions, with the canonical basis providing better results up until a bond length of \toadd{$\sim 2.5\,a_0$}, at which point the local basis becomes more accurate.
At short bond lengths, the canonical basis allows a description of the dominant kinetic energy driven effects with a single configuration, while the `Mott insulating' stretched geometries which are dominated by the interactions favor the local basis as an efficient representation due to the rapidly decaying correlation lengths.
Nevertheless, the geometry change coupled to fermionic antisymmetry necessitates a strongly signed set of wave function amplitudes, requiring complex parameters.
Furthermore, the ambiguity in defining an ordering of the orbitals (in both representations) impacts upon the accuracy that can be achieved in the state, significantly increasing the relative energy error compared to the 1D system.

\section{Conclusions and outlook}

With the emergence of highly expressive functional models based on machine learning paradigms as ansätze for the many-body quantum state, autoregressive models seem particularly appealing due to their inherent design which allows for an exact generation of configurational samples.
We have presented a general framework for the construction of these autoregressive forms from general approximators, defining the two constraints which must be imposed on their form in terms of masking and normalization steps.
Exemplifying the construction for the recently introduced Gaussian process state for the conditional probability distributions which make up these models, we introduced a new autoregressive ansatz, explicitly underpinned by physical modelling assumptions which motivate the GPS ansatz, and adapted for autoregressive sampling.
Furthermore, we go beyond autoregressive adaptations of quantum states to consider `filters', designed to model correlations in a translationally symmetric fashion, and allow for a corresponding scaling reduction in parameter numbers.
We show how these can then be combined with the quantum states in both a general framework, and specifically with the GPS.

While the benefits of direct sampling have been previously highlighted, with the practical optimization difficulties of these expressive states well known~\cite{szaboNeuralNetworkWave2020,westerhoutGeneralizationPropertiesNeural2020, bukovLearningGroundState2021,kochkovVariationalOptimizationAI2018,rothHighaccuracyVariationalMonte2023,chenEfficientOptimizationDeep2023}, we put particular focus on the ramifications for the variational flexibility of these \toadd{modified} states due to these autoregressive and filter adaptations compared to their parent \toadd{GPS} model.
This was numerically investigated for the variational optimization of unknown ground states across spin, fermionic and {\em ab initio} systems, highlighting that for the benefits and simplicity of direct sampling of a normalized autoregressive quantum state there can be a significant loss in expressibility.
We numerically investigate which of the two constraints (masking or normalization) required for an autoregressive state this primarily stems from, finding (perhaps surprisingly) that the explicit normalization affects the expressibility more than the masking constraint of an ordered and causal set of conditionals.
While numerical results were specifically obtained from the simple (yet nonetheless universal) autoregressive GPS model, we believe that these general conclusions would transfer to other forms of flexible ansatz, with the choice of `parent' architecture for the conditionals less important in the flexibility of these states compared to the underlying assumptions required for the autoregressive property to emerge. \toadd{We did not balance this loss of flexibility with the benefits of direct sampling in terms of an overall picture of the net benefit of autoregressive model. This is likely to be highly system-dependent, varying with the ease and ergodicity in the sampling of the configurational space, making broad conclusions impossible. Nevertheless, it should be cautioned that there is potentially a price in flexibility to pay when using autoregressive models compared to their parent model in systems where the traditional configurational sampling is not difficult, in particular with improved sampling schemes emerging~\cite{bravyiRapidlyMixingMarkov2023}.}
We show that the autoregressive model especially performs well when capturing the correlations emerging from (quasi-)one dimensional systems, where a natural order for the decomposition into a product of conditionals can be found.
Indeed, we are able to demonstrate a high degree of accuracy for one-dimensional fermionic systems within different settings and correlation regimes, here exemplified for prototypical Hubbard models as well as fully {\em ab initio} descriptions of the electronic structure of hydrogen atom arrays.
We furthermore compare the performance across signed and unsigned states, as well as the importance of basis choice in moving towards {\em ab initio} systems.
Generalizing these constructions for models in higher dimensions, as has started to be done for e.g. recurrent neural networks \cite{hibat-allahRecurrentNeuralNetwork2020,wuTensornetworkQuantumStates2023}, is an ongoing direction of future work.
Bringing these different modelling paradigms together in a general framework can build us towards a practical tool for the description of general quantum states, from their variational optimization, to time-evolution~\cite{carleoSolvingQuantumManybody2017,hofmannRoleStochasticNoise2022,linScalingNeuralNetworkQuantum2022,donatellaDynamicsAutoregressiveNeural2023}
or the simulation of quantum circuits~\cite{jonssonNeuralnetworkStatesClassical2018,medvidovicClassicalVariationalSimulation2021}.

\section*{Code Availability}
The code for this project was developed as part of the \href{https://github.com/BoothGroup/GPSKet}{GPSKet} plugin for \href{https://github.com/netket/netket}{NetKet}~\cite{carleoNetKetMachineLearning2019,vicentiniNetKetMachineLearning2022} and is made available, together with configurations files to reproduce the figures in the paper, at \url{https://github.com/BoothGroup/GPSKet/tree/master/scripts/ARGPS}.

\begin{acknowledgments}
    The authors gratefully acknowledge support from the Air Force Office of Scientific Research under award number FA8655-22-1-7011, as well as the European Union’s Horizon 2020 research and innovation programme under grant agreement No. 759063. We are grateful to the UK Materials and Molecular Modelling Hub for computational resources, which is partially funded by EPSRC (EP/P020194/1 and EP/T022213/1). Furthermore, we acknowledge the use of the high performance computing environment CREATE at King’s College London~\cite{kingscollegelondone-researchteamKingComputationalResearch2022}.
\end{acknowledgments}

\bibliographystyle{plainnat}

\appendix

\section{Optimization Details}\label{app:optimization}
Throughout this work we optimize all the ansätze with an improved Stochastic Reconfiguration (SR)~\cite{sorellaGeneralizedLanczosAlgorithm2001} algorithm introduced in Ref.~\onlinecite{lovatoHiddennucleonsNeuralnetworkQuantum2022}, which we implemented in our GPSKet plugin for NetKet~\cite{carleoNetKetMachineLearning2019,vicentiniNetKetMachineLearning2022}.
In the SR scheme, parameters are updated according to the following rule:
\begin{equation}\label{eq:sr-update}
    \theta_{t+1} = \theta_t - \eta S^{-1} g,
\end{equation}
where $\eta$ is the step size (learning rate), $\theta_t$ are the parameters of the ansatz at iteration $t$, $S$ is the quantum geometric tensor (QGT) and $g$ is the variational energy gradient.

The QGT and the energy gradient, can be defined by introducing operators $\hat{O}_k$ representing the derivative with respect to the $k$-th parameter of the log wave function amplitude according to
\begin{equation}
    \braket{\mathbf{x}|\hat{O}_k|\mathbf{x}'} = \delta_{\mathbf{x},\mathbf{x}'}\frac{\partial \log{\psi_{\theta}(\mathbf{x})}}{\partial \theta_k},
\end{equation}
where $\ket{\mathbf{x}}$ and $\ket{\mathbf{x}'}$ are computational basis states.
The QGT and the energy gradient can then be evaluated via Monte Carlo sampling of the following expectation values:
\begin{align}
    S_{i,j} &= \langle\hat{O}^*_i\hat{O}_i\rangle - \langle\hat{O}^*_i\rangle\langle\hat{O}_i\rangle, \\
    g_i &= \langle\hat{O}^*_i\hat{H}\rangle - \langle\hat{H}\rangle\langle\hat{O}_i\rangle.
\end{align}

It is typically required to appropriately regularize the solution of the update of Eq.~\eqref{eq:sr-update}, which involves solving for the update vector $S^{-1} g$.
A common strategy is to add a constant shift to the diagonal of the $S$ matrix.
Instead of applying a constant shift to the diagonal of the $S$ matrix to stabilize its inversion in Eq. \ref{eq:sr-update}, we update the diagonal entries of $S$ with a parameter-dependent shift based on the scheme introduced in Ref.~\onlinecite{lovatoHiddennucleonsNeuralnetworkQuantum2022}.
We found that this approach sometimes significantly helps to reliably optimize the autoregressive parameterization.
The scheme is based on adding a regularization shift to the diagonal of the $S$ matrix based on the exponential moving average of the squared gradient, $v_t$, effectively rotating the parameter updates towards the RMSProp gradient descent update directions~\cite{hintongeoffreyLecture6aOverview2012}.
This means that the $S$ matrix is regularized by replacing it according to
\begin{equation}
    S \mapsto (1-\varepsilon)S + \varepsilon\text{diag}(\sqrt{v}+10^{-8}),
\end{equation}
which depends on an additional hyperparameter $\varepsilon$ between $0$ and $1$, controlling the amount of regularization.
The exponentially moving average of squared gradients is continuously updated over the course of the optimization according to
\begin{equation}
    v \mapsto \beta v' + (1-\beta)g^2,
\end{equation}
where $v'$ is the accumulated value from the previous iteration, and an additional momentum hyperparameter $\beta$ controls the rate of the decay.

Within all our numerical tests, we set the momentum value to $\beta = 0.9$.
We chose a learning rate of $\eta=0.01$ with a diagonal shift constant of $\varepsilon=0.1$ for our simulation with lattice models, and a learning rate of $\eta=0.04$ with a shift constant $\varepsilon=0.01$ for the {\em ab initio} simulations of hydrogen systems.
We computed estimates of the variational energy, the gradient, and $S$ matrix elements with 4096 (non-symmetric representations), or 1024 (symmetric representations) for lattice models, and with $5000$ samples for the {\em ab initio} systems.
For non-autoregressive models, we relied on the Metropolis-Hastings algorithm based on spin exchange proposals to generate samples according to the Born distribution defined by the ansatz.
The reported final energies were computed by averaging the sampled variational energy over the last 50 iterations.

\section{Representing product states with autoregressive GPS}\label{app:ar-product-states}
While the practical applications studied in this work specifically focus on capturing non-trivial correlations between the modes with the machine learning inspired ansatz, the model should also be able to reproduce physical characteristics of non-entangled states, as, e.g., obtained for eigenstates of Hamiltonians with vanishing couplings between system fragments.
In particular the ability to represent such simple product states with the model is likely an important building block to model ground states typically displaying a low, but non-vanishing, degree of entanglement~\cite{eisertAreaLawsEntanglement2010a}.
In this appendix, we show how these unentangled states can also be obtained with the autoregressive extensions of the GPS model considered in the main text.

A general product state for a system comprising $N$ modes decomposes as
\begin{equation}
    \ket{\psi} = \otimes_{i=i}^N\ket{\psi_i},
\end{equation}
where the states $\ket{\psi_i}$ are states only associated with the local Hilbert space of the $i$-th mode.
This means that wave function amplitudes of the configurations in the computational basis for this state evaluate to
\begin{equation}
    \psi(\mathbf{x}) = \prod_{i=1}^N c^i_{x_i},
\end{equation}
with an $N \times D$ tensor of local amplitudes $c^i_{x_i} = \braket{x_i | \psi_i}$.

To represent a general product state by an autoregressive model, we decompose the wave function amplitudes according to
\begin{equation}
    \psi_{AR}(\mathbf{x}) = \prod_{i=1}^N\frac{\tilde{\psi}_i(x_{i}|\mathbf{x}_{<i})}{\sqrt{\sum\limits_{x'=0}\limits^{D-1}|\tilde{\psi}_i(x'|\mathbf{x}_{<i})|^2}},
\end{equation}
we represent the local amplitudes $c^i_{x_i}$ by the conditional wave functions amplitudes $\tilde{\psi}_i(x_{i}|\mathbf{x}_{<i})$.

It can directly be seen that the general autoregressive GPS model as defined in Eq.~\ref{eq:argps-ansatz} of the main text, which specifies the wave function amplitudes as
\begin{equation}
    \psi_{AR-GPS}(\mathbf{x}) = \prod_{i=1}^N\frac{\exp{\left(\sum\limits_{m=1}\limits^M\prod\limits_{j\leq i}\epsilon_{x_j,m,j}^{(i)}\right)}}{\sqrt{\sum\limits_{x'=0}\limits^{D-1}\left|\exp{\left(\sum\limits_{m=1}\limits^M\epsilon_{x',m,i}^{(i)}\prod\limits_{j<i}\epsilon_{x_j,m,j}^{(i)}\right)}\right|^2}},
\end{equation}
can represent arbitrary product states with a support dimension $M=1$, by employing the following choice
\begin{equation}
    \epsilon_{x_j,m,j}^{(i)} = \begin{cases} \log(c^i_{x_i}) \quad &\text{if } j = i\\ 1 \quad &\text{otherwise} \end{cases}.
\end{equation}

As an approach to impose additional structure into the ansatz (and reduce the number of variational parameters), we introduced filter-based version of (autoregressive) GPS models.
This relies on transferring a symmetric structure of the system to the model similar to that of a convolutional neural network, and is compatible with the autoregressive adaptation, since an additional masking can always be applied in order to ensure the autoregressive property is maintained.
Applying this to the product state representation above, results in a fully symmetric product state where all the local states $\ket{\psi_i}$ are equal, i.e., a wave function decomposing as a product with mode-independent amplitudes $c^i_{x_i}$ according to
\begin{equation}
    \psi(\mathbf{x}) = \prod_{i=1}^N c_{x_i}.
\end{equation}
While this filtering approach reduces the number of variational parameters (thus often improving the practical optimizability of the state) for a given support dimension, the fully-symmetric product state representation can only be sensible if the target agrees with this trivial symmetry.

As an alternative to the filtering approach to impose additional structure, in the main text we also consider a `weight sharing' approach in which parameters are equivalent among the conditionals of different sites according to the model
\begin{equation}
    \psi_{AR-GPS}(\mathbf{x}) = \prod_{i=1}^N\frac{\exp{\left(\sum\limits_{m=1}\limits^M\prod\limits_{j\leq i}\epsilon_{x_j,m,j}\right)}}{\sqrt{\sum\limits_{x'=0}\limits^{D-1}\left|\exp{\left(\sum\limits_{m=1}\limits^M\epsilon_{x',m,i}\prod\limits_{j<i}\epsilon_{x_j,m,j}\right)}\right|^2}},
\end{equation}
characterized by $M \times N \times D$ parameters $\epsilon_{x,m,j}$.
This ansatz has a factor $\mathcal{O}(N)$ fewer parameters, and caching intermediate values of the product over sites $j$ allows for a reduction of the computational cost which is linear in the system size when sampling and evaluating configurations.

However, with these additional weight sharing constraints on the model, it is no longer obvious how to represent arbitrary product states most compactly, let alone with a constant support dimension $M=1$, since the same parameters are used in all the correlators.
We can still recover a representation of arbitrary product states by using a support dimension matching the size of the system, $M=N$, in which case a representation arbitrary product states with an autoregressive weight-sharing ansatz can be obtained by choosing the model parameters as
\begin{equation}
    \epsilon_{x_j,m,j} = \begin{cases} \log(c^j_{x_j}) \quad &\text{if } j = m \\ 1 &\text{if } j < m  \\  0 \quad &\text{otherwise} \end{cases}.
\end{equation}
The required increase in the support dimension of the model to represent fully unentangled states therefore suggests that a weight-sharing construction might not be as suitable to target states exhibiting low degrees of entanglement, representing a major drawback of such a construction.
This is also in agreement with results from numerical experiments, where we commonly observed a significant decay of the achievable accuracy when utilizing the autoregressive ansatz based on a weight-sharing parameter reduction. 

\section{Fast updating of the AR-GPS} \label{app:fastupdate}
To reduce the overall computational scaling, it is often useful to exploit the fact that the evaluation of local energies generally requires low-rank updates to wave function amplitudes arising from few-electron changes to configurations of interest.
These are applicable for $k$-local Hamiltonians which connect a sampled configuration with other computational basis states that only differ in their occupancy for few sites.
While this is, in general, true for all the systems considered in this work, the utilization of a fast updating strategy to evaluate the amplitude of the connected configurations typically becomes particularly important for the {\em ab initio} Hamiltonians where each basis state is connected to a quartically-scaling number of connected configurations.
In this section, we show how updates to wave function amplitudes can be implemented for the AR-GPS model, resulting in an $\mathcal{O}(N)$ scaling improvement of the connected amplitude evaluations. 

The fast updating scheme directly follows from the approach for GPS amplitudes as also outlined in Ref.~\cite{rathFrameworkEfficientInitio2023}.
With the definition of the AR-GPS ansatz according to Eq.~\eqref{eq:argps-ansatz}, the model associates a wave function amplitude with a basis configuration $\mathbf{x}$ according to
\begin{equation}
    \psi_{AR}(\mathbf{x}) = \prod_{i=1}^N\frac{\exp{\left(\sum\limits_{m=1}\limits^M\varphi_{i,m}(\mathbf{x})\right)}}{\sqrt{\sum\limits_{x'=0}\limits^{D-1}\left|\exp{\left(\sum\limits\limits_{m=1}\limits^M \epsilon_{x',m,i}^{(i)} \, \varphi_{i,m}(\mathbf{x})/\epsilon_{x_i,m,i}^{(i)} \right)}\right|^2}}.
\end{equation}
Here, we introduced the $N \times M$ products $\varphi_{i,m}$, which are defined as
\begin{equation}
    \varphi_{i,m}(\mathbf{x}) = \prod_{j \leq i}\epsilon_{x_j,m,j}^{(i)}.
\end{equation}
The direct evaluation of an amplitude is therefore associated with a cost of $\mathcal{O}(N^2 M)$.

To avoid redundant computations of elements for an update of the amplitude for a connected configuration $\tilde{\mathbf{x}}$ with a similar occupancy as the initial configuration $\mathbf{x}$, we can consider updates to values of the products $\varphi_{i,m}$.
Caching these for configuration $\mathbf{x}$, its value can be updated for a configuration $\tilde{\mathbf{x}}$ according to
\begin{equation}
    \varphi_{i,m}(\tilde{\mathbf{x}}) = \varphi_{i,m}(\mathbf{x}) \times \prod_k \theta^{(k)}_{i,m}(x_k, \tilde{x}_k),
\end{equation}
where the product only runs over those site indices for which the occupancy is different in configurations $\mathbf{x}$ and $\tilde{\mathbf{x}}$.
The update factor $\theta^{(k)}_{i,m}(x_k, \tilde{x}_k)$ is given as
\begin{equation}
    \theta^{(k)}_{i,m}(x_k, \tilde{x}_k) = \begin{cases} \epsilon_{\tilde{x}_k,m,k}^{(i)}/\epsilon_{x_k,m,k}^{(i)} \quad \text{if } k \leq i \\ 1 \quad \text{otherwise} \end{cases},
\end{equation}
and can therefore easily be evaluated in constant time.
This means that the full cost to update the amplitude for the connected configuration $\tilde{\mathbf{x}}$ only scales as $\mathcal{O}(N M K)$, where $K$ is the number of local updates which are employed.
Within the considered lattice models with nearest neighbor interactions, the number of updates is at most $K=2$, and for {\em ab initio} systems the occupancy changes on at most $4$ orbitals through the application of the Hamiltonian.

\end{document}